\documentclass[a4paper,
               keeplastbox,   % flushend option: not to un-indent last line in References
               ]{jacow}
\usepackage{pdfpages,multirow,ragged2e} %
\usepackage{comment}
\makeatletter%
	\ifboolexpr{bool{xetex}}
	 {\renewcommand{\Gin@extensions}{.pdf,%
	                    .png,.jpg,.bmp,.pict,.tif,.psd,.mac,.sga,.tga,.gif,%
	                    .eps,.ps,%
	                    }}{}
\makeatother

\ifboolexpr{bool{xetex} or bool{luatex}} % test for XeTeX/LuaTeX
 {}                                      % input encoding is utf8 by default
 {\usepackage[utf8]{inputenc}}           % switch to utf8

\usepackage[USenglish]{babel}

\ifboolexpr{bool{jacowbiblatex}}%
 {%
  \addbibresource{jacow-test.bib}
  \addbibresource{biblatex-examples.bib}
 }{}
\begin{document}

\title{Measures to mitigate the coherent beam-beam instability at \NoCaseChange{CEPC}\thanks{\NoCaseChange{Work supported by the innovation study of IHEP and the Chinese Academy of Sciences President’s International Fellowship Initiative (Grant No. 2024PVA0057).}}}

\author{Y. Zhang\thanks{zhangy@ihep.ac.cn}, N. Wang, L. Pan, IHEP, Beijing, China\\
C. Lin, IASF, Shenzhen, China \\
K. Ohmi$^1$, KEK, Tsukuba, Japan \\
  $^1$ Also at IHEP, Beijing, China}
\maketitle

\begin{abstract}
Both horizontal and vertical coherent beam-beam instability are important issues at CEPC. The horizontal instability (X-Z instability) could be induced by beam-beam itself. In this paper we try to study the effect of chromaticity and resistive feedback by analysis and simulation. The vertical instability may be induced due to the combined effect of beam-beam interaction and vacuum impedance. Finite chromaticity and asymmetrical tunes have been proposed to suppress the vertical instability. Due to the further increase of impedance budget, we need to find more measures to mitigate the instability. The effect of resistive feedback and hourglass effect are evaluated by simulation. 
\end{abstract}

\section{INTRODUCTION}

After CEPC was proposed near 2012, the conceptual design report (CDR) \cite{cdr} and technical design report (TDR) \cite{tdr} has been published respectively in 2018 and 2023.
The CEPC has two interaction points (IPs) for e+e– collisions and is designed to operate in four energy modes (ttbar, Higgs, W, and Z).
We only focus on the Z mode in this paper, since the collective effect is most serious \cite{wangna-jinst-2024}.

%In order to reach high luminosity, CEPC are expected to use the crab waist collision scheme [cite] with a large Piwinski angle.
%The CW scheme has also become the baseline design of SuperB, SuperCT and FCCee [4–8, prab].

The crab-waist scheme \cite{mikhail-prl} has been the baseline of circular e+e- colliders.
During the design of new-era storage ring e+e- colliders, a coherent beam-beam head-tail instability (X-Z instability) has been revealed via beam-beam simulation and later validated by the mode analysis method \cite{ohmi-prl-2018,kuroo-prab-2018}. This instability sets a new constraint on machine parameters and has greatly impacted the design of CEPC and FCCee.

The combined effects between beam-beam interaction and longitudinal impedance have been studied by numerical simulations and analysis method \cite{zhang-prab-2020,lin-prab-2022}. For CEPC CDR parameters, it has been found that there would not exist a large enough stable tune area when the effect of longitudinal impedance is considered. These findings indicate that in the present and future colliders with crab waist and large crossing angles, different dynamic effects should be modeled self-consistently in simulations and/or analyses to ensure a reliable prediction of beam stability.

Recently,
strong-strong beam-beam simulations with vertical impedance showed that a $\sigma$-mode head-tail instability may appear in CEPC and SuperKEKB \cite{zhang-prab-2023,zhou-prab-2023}.
The results showed that this vertical instability can appear below the transverse mode-coupling instability (TMCI) threshold (here, the TMCI threshold is determined by vertical impedance without collision). 
Analysis shows that \cite{zhang-prab-2023,ohmi-prab-2023} the 0 mode tune would decrease with bunch population due to ring impedance, while the -1 mode tune would increase due to cross-wake impedance.
That is to say, the TMCI threshold would reduce when both beam-beam interaction and ring impedance are considered.

The impedance model of CEPC has been continuously updated \cite{wangna-jinst-2024} with the evolvement of design work.
It seems that impedance budget would increase with more detailed components are included.
It is very important to evaluate and mitigate the collective instability with collision.
The paper is organized as follows: the machine parameters and impedance are first introduced. The effect of chromaticity and resistive feedback on single bunch instability are then discussed. Different mitigation methods on horizontal and vertical beam-beam instability are respectively studied by simulation and/or analysis. In the end, the work is summarized and outlooked.

\section{PARAMETER LIST AND IMPEDANCE MODELING}

The main parameters used in the paper are listed in Table~\ref{tab:parameter}, which are the Z mode parameters in the TDR.
Since there are two symmetrical IPs in the rings, the parameters listed here are for half ring.

\begin{table}[!hbt]
   \centering
   \caption{Main Parameters}
   \begin{tabular}{lc}
       \toprule
       \textbf{Parameter} & \textbf{Value} \\
       \midrule
Beam energy (GeV) & 45.5 \\
Bunch population ($10^{10}$) & 14 \\
Momentum compaction factor ($10^{-5}$) & 1.43 \\ 
Transverse damping time (turns) & 4919 \\
Synchrotron tune & 0.0176 \\
Emittance Hor (nm)/Vert (pm) & 0.27/1.4 \\
Bunch length (mm) (SR/BS)$^*$ & 2.5/8.7 \\
Energy spread (\%) (SR/BS)$^*$ & 0.04/0.13  \\
Piwinski angle (BS)$^*$ & 24.2  \\
$\xi_x / \xi_y$ & 0.004/0.127 \\
$\beta^*$ Hor (m)/Vert (mm) & 0.13/0.9  \\
Luminosity/IP ($10^{34}$/cm$^{2}$s) & 115  \\
     \bottomrule
   \end{tabular} \\
$^*$SR: synchrotron radiation, BS: beamstrahlung 
   \label{tab:parameter}
\end{table}

Here we only present a very brief introduction on the impedance modeling. More details are described in \cite{wangna-jinst-2024}.
The resistive wall impedance, generated by the finite conductivity of the vacuum chambers, is one of
the dominant impedance contributors. The main vacuum chamber is made of copper with a layer
of 0.2 $\mu$m non-evaporable getter (NEG) coating on its inner surface.
The geometrical impedance generated by the discontinuity of the vacuum chambers is calculated
individually assuming that there is no cross talk between the adjacent components. The impedance
is calculated with the numerical code CST Particle Studio \cite{CST} and ABCI \cite{ABCI}. The total impedance
for the ring is a sum of the individual contributors. The geometrical components considered in the
impedance modelling include RF cavities, flanges, bellows, gate valves, vacuum pumps, BPMs, IR
collimators, electro separators, IP chambers, and vacuum transitions.

\section{SINGLE BUNCH INSTABILITY}

The bunch population of TMCI threshold is about $21\times 10^{10}$, where the equilibrium bunch length with collision is used.
However the threshold reduces to about $15\times 10^{10}$ with new impedance budget other than TDR version.
With the updated impedance, the TMCI threshold is too close to the design bunch population.

\begin{figure}[!htb]
   \centering
   \includegraphics[width=.9\columnwidth]{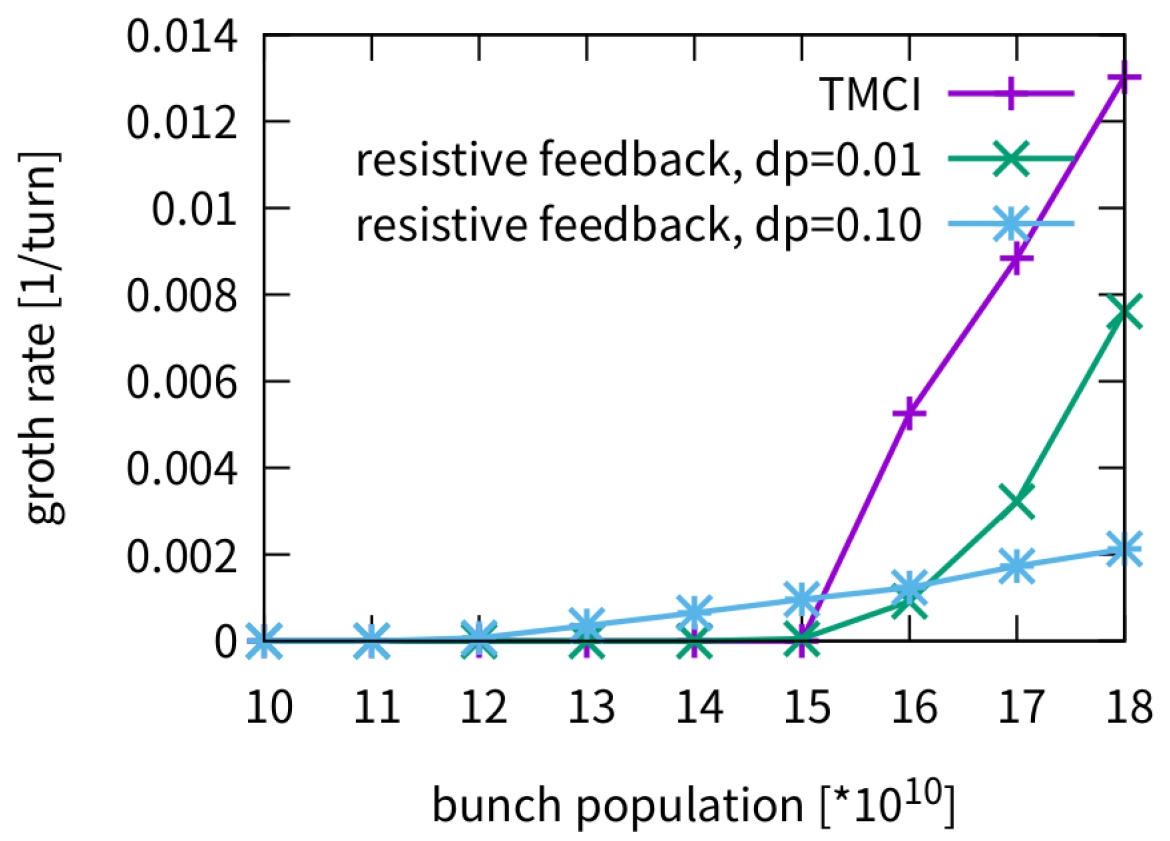}
   \caption{Effect of resistive feedback on TMCI (w/o collision).}
   \label{fig:growth-tmci-feedback-wocol}
\end{figure}

Here we first check the effect of resistive feedback on the single bunch instability.
In the simulation, the resistive feedback is modeled by
\begin{equation}
    \delta p = -2d_p \langle p \rangle
\end{equation}
where $d_p$ is the damping decrement.
In Fig.~\ref{fig:growth-tmci-feedback-wocol} the growth rate of TMCI versus bunch population with different feedback strength is shown. It could be concluded that the growth rate is reduced with finite resistive feedback.
However the instability threshold is also lower with resistive feedback.

This phenomenon has been studied in \cite{elias-itsr}, where it is found that the lower threshold than TMCI instability is not due to mode coupling and the instability is called ``Imaginary tune split and repulsion single-bunch instability'' (ITSR).
More recently it has also been found that `-1' mode instability \cite{ohmi-minus1mode} may be induced by feedback at LER ring of SuperKEKB. 

\begin{figure}[!htb]
   \centering
   \includegraphics[width=.9\columnwidth]{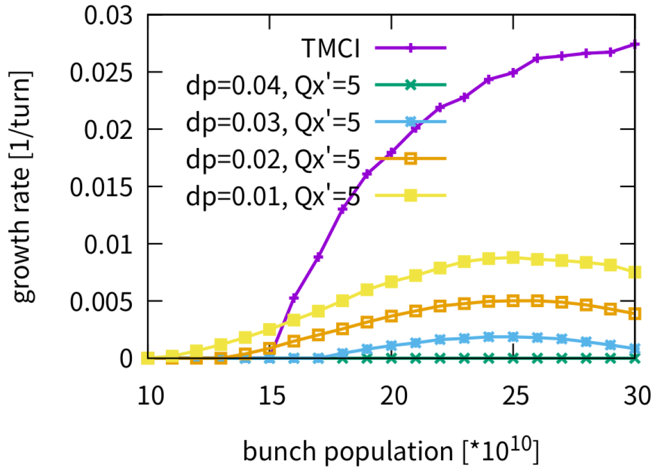}
   \caption{Effect of resistive feedback on TMCI (w/o collision).}
   \label{fig:growth-tmci-feedback-wocol-qxp5}
\end{figure}

The above results are obtained with tune chromaticity $Q_x'=0$.
We also try finite chromaticity ($Q_x'=5$) with different feedback strength.
The results are shown in Fig.~\ref{fig:growth-tmci-feedback-wocol-qxp5}. 
It seems the single bunch instability could be mitigated by the combination of a strong resistive feedback and finite tune chromaticity.
One should be reminded that a very strong bunch-by-bunch feedback system are required to suppress the coupled bunch instability at CEPC, where the instability growth time could be about 2~ms \cite{wangna-jinst-2024}.

\section{HORIZONTAL BEAM-BEAM INSTABILITY}
The horizontal head-tail beam-beam instability (X-Z instability) was induced by the localized property of beam-beam interaction.
There would not exist horizontal dipole oscillation in the instability, while there would exist growth of skew moment $\langle xz \rangle$ and horizontal emittance.
It has been indicated by D. Shatilov that feedback could not suppress this instability \cite{shatilov-icfa}. Our simulation also confirm this comment.

\begin{figure}[!htb]
   \centering
   \includegraphics[width=.9\columnwidth]{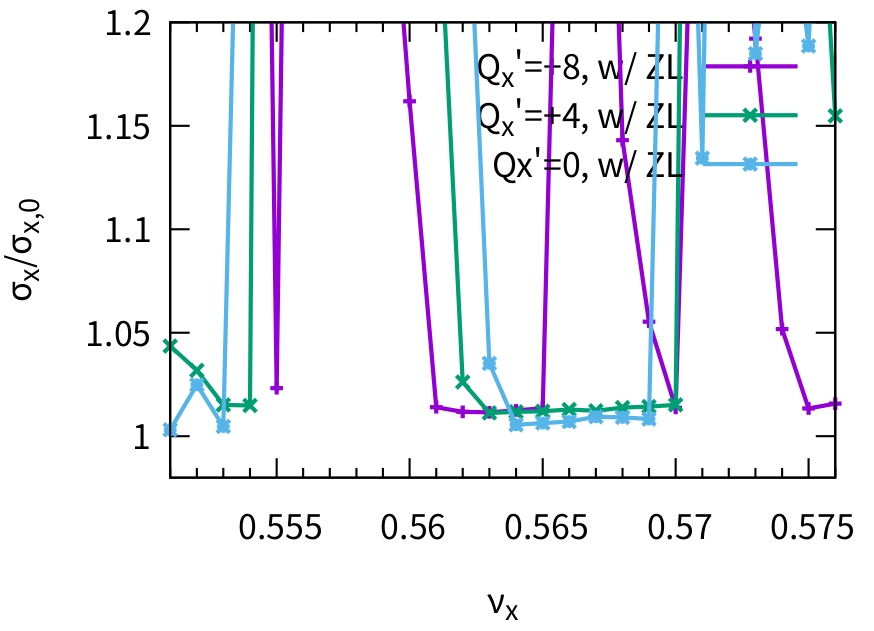}
   \caption{Horizontal beam size blowup versus horizontal tune with finite tune chromaticity and longitudinal impedance.}
   \label{fig:rmsx-vs-nux-xix-zl}
\end{figure}

Chromaticity is scanned with and without longitudinal impedance.
It is found that linear chromaticity is detrimental for the X-Z instability when longitudinal impedance is not considered.
Figure~\ref{fig:rmsx-vs-nux-xix-zl} shows that proper tune chromaticity ($Q_x'=4$) may enlarge the stable tune area when potential-well distrotion effect is included. However large tune chromaticity ($Q_x'=8$) would be harmful for the stability.

\begin{figure}[!htb]
   \centering
   \includegraphics[width=.9\columnwidth]{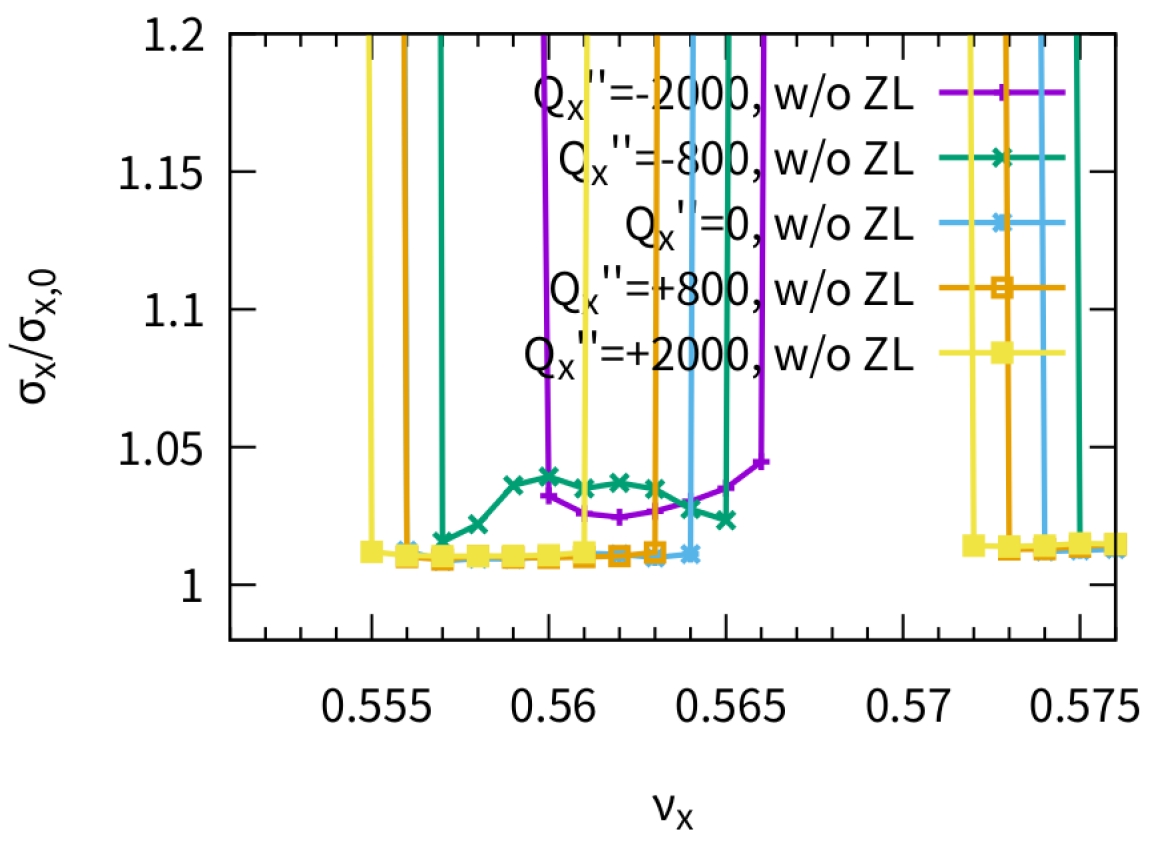}
   \caption{Horizontal beam size blowup versus horizontal tune with 2nd order tune chromaticity.}
   \label{fig:rmsx-vs-nux-xix2-wozl}
\end{figure}

The 2nd order chromaticity is also scanned without longitudinal impedance.
The results are shown in Fig.~\ref{fig:rmsx-vs-nux-xix2-wozl}. It could be concluded that the 2nd order chromaticity is detrimental for collision stability. 
The left/right side of stable tune region may become unstable for minus/positive $Q_x''$.
The phenomenon agree with the analysis results.

No stable working points are found when we consider transverse impedance.
Even the feedback system could not help suppress X-Z instability, it could help mitigate the single bunch instability which is induced by transverse ring impedance.
The collision with both transverse/longitudinal impedance, strong resistive feedback ($d_p=0.05$) and finite chromaitcity ($Q_x'=5$) shows that the stable tune region is same as that of collision without transverse impedance.

\section{VERTICAL BEAM-BEAM INSTABILITY}

It has been found that the beam-beam interaction would induce the increase of -1 mode (azimuthal) tune with bunch population \cite{zhang-prab-2023,ohmi-prab-2023}.
It is well know that the conventional ring impedance would induce the reduction of 0 mode (azimuthal) tune with bunch population.
Then the instability threshold of vertical TMCI in collision would be lower that of single bunch due to the interplay of beam-beam and ring impedance.

Finite tune chromaticity is helpful to suppress the vertical coherent beam-beam instability \cite{zhang-prab-2023}.
Since the beam-beam instability is $\sigma$ mode, the asymmetry of colliding bunch working points also help mitigate the instability \cite{zhang-prab-2023}, which is similar in the horizontal direction.

After the impedance is updated, the conventional TMCI threshold is about $15\times 10^{10}$ without collision. 
It is found that only $Q_y'=10$ could not mitigate the vertical beam-beam instability.
Together with finite chromaticity, a strong resistive feedback $d_p=0.05$ succeed in suppressing the instability.

\begin{figure}[!htb]
   \centering
   \includegraphics[width=.9\columnwidth]{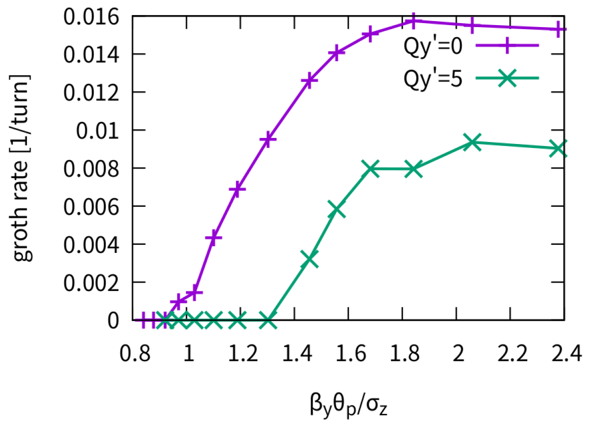}
   \caption{Growth rate versus the ratio between $\beta_y^*$ and overlapping area $\sigma_z/\theta_P$. Here the bunch length is assumed as that of design equilibrium. Longitudinal impedance is considered.}
   \label{fig:growth-versus-hourglass-effect}
\end{figure}

With the large Piwinski angle $\theta_P=\sigma_z\theta/\sigma_x$, the ratio between $\beta_y^*$ and overlapping area is $\beta_y^*\theta_P/\sigma_z$.
Just as a try, we change the horizontal emittance while keeping other parameters unchanged to study the hourlgass effect.
The beamstrahlung effect is ignored, and the bunch length is assumed as that of equilibrium $\sigma_z$ in collision with beamstrahlung.
Figure~\ref{fig:growth-versus-hourglass-effect} shows that the instability could be suppressed if $\beta_y^*\theta_P/\sigma_z<1.3$ with $Q_y'=5$.
The hourglass effect has also been studied and found helpful in \cite{ohmi-prab-2023}.

\section{SUMMARY}

This paper introduce the challenges on coherent beam-beam instability at CEPC.
We studied a few measures to mitigate the instabilities.
For single bunch (w/o collision), strong resistive feedback together with finite chromaticity could help mitigate the TMCI.
Linear tune chromaticity could help mitigate horizontal beam-beam instability only when the PWD is considered.
Strong feedback does not help mitigate the X-Z instability, however it help mitigate the TMCI and further ensure the collision stability.
Tune asymmetry could help mitigate both horizontal and vertical instability since $\sigma$ mode dominates.
Vertical tune chromaticity and feedback also help mitigate the vertical instability (TMCI) in collision.
Hourglass effect may also help, however machine parameters need optimization to make it work.
With the new developed tool \cite{li-nima-2024}, the effect of full lattice on the instability will be studied. 
This is an ongoing work, further simulation and analysis need be done, as well as the more realistic feedback should be considered.

\section{ACKNOWLEDGEMENTS}
The first author would like to thank M. Zobov, M. Migliorati, K. Oide, D. Shatilov and D. Zhou for fruitful discussions.

\end{document}